\documentclass[12pt]{article}
\usepackage{latexsym}
\usepackage{amsmath,,calrsfs}
\usepackage{amsfonts}
\usepackage{amssymb}
\usepackage{amscd}
\usepackage{bbm}
\usepackage{fancybox}
\usepackage{cite}
\usepackage{amsmath,amsfonts,amsbsy}
\usepackage{pstricks,pst-node}
\usepackage[small,bf,hang]{caption2}
\usepackage{graphicx}
\usepackage{epsfig}
\usepackage{psfrag}
\usepackage{comment}
\usepackage{hyperref}

\usepackage{float}

\psset{unit=1.3cm,linewidth=.5pt,radius=.2}  

\usepackage{multirow}                     
\usepackage{float}                          
\usepackage{lscape}                         
\usepackage{bm}

\newcommand{\cL}{{\cal L}}

\newcommand{\beq}{\begin{equation}}
\newcommand{\eeq}{\end{equation}}
\newcommand{\bi}{\begin{itemize}}
\newcommand{\ei}{\end{itemize}}
\newcommand{\bt}{\begin{tabular}}
\newcommand{\et}{\end{tabular}}
\newcommand{\bc}{\begin{center}}
\newcommand{\ec}{\end{center}}

\newcommand{\be}{\begin{equation}}
\newcommand{\ee}{\end{equation}}
\newcommand{\bea}{\begin{eqnarray}}
\newcommand{\eea}{\end{eqnarray}}
\newcommand{\ba}{\begin{array}}
\newcommand{\ea}{\end{array}}

\def\bbox{{\,\lower0.9pt\vbox{\hrule \hbox{\vrule height 0.2 cm
\hskip 0.2 cm \vrule height 0.2 cm}\hrule}\,}}
\newcommand{\dsl}{\pa \kern-0.5em /}

\def\tr{{\rm tr}}

\def\tr{{\rm tr}}

\makeatletter \@addtoreset{equation}{section} \makeatother

\def\slashchar#1{\setbox0=\hbox{$#1$}           
   \dimen0=\wd0                                 
   \setbox1=\hbox{/} \dimen1=\wd1               
   \ifdim\dimen0>\dimen1                        
      \rlap{\hbox to \dimen0{\hfil/\hfil}}      
      #1                                        
   \else                                        
      \rlap{\hbox to \dimen1{\hfil$#1$\hfil}}   
      /                                         
   \fi}

\begin{document}

\begin{titlepage}
\begin{center}

\vskip 1.5cm

{\Large \bf On-shell {\it versus} Off-shell  Equivalence in 3D Gravity}

\vskip 1cm

{\bf Eric A. Bergshoeff\,${}^1$,  Wout Merbis\,${}^2$ and
Paul K.~Townsend\,${}^3$} \\

\vskip 25pt

{\em $^1$  \hskip -.1truecm
\em Van Swinderen Institute, University of Groningen, \\ Nijenborgh 4,
9747 AG Groningen, The Netherlands
 \vskip 5pt }

{email: {\tt  E.A.Bergshoeff@rug.nl}} \\

\vskip .4truecm

{\em $^2$ \hskip -.1truecm
\em  Universit\'e Libre de Bruxelles and International Solvay Institutes,\\ Physique Th\'eorique et Math\'ematique,
Campus Plaine - CP 231, \\
B-1050 Bruxelles, Belgium \vskip 5pt }

{email: {\tt  wmerbis@ulb.ac.be}} \\

\vskip .4truecm

{\em $^3$ \hskip -.1truecm
\em  Department of Applied Mathematics and Theoretical Physics,\\ Centre for Mathematical Sciences, University of Cambridge,\\
Wilberforce Road, Cambridge, CB3 0WA, U.K.\vskip 5pt }

{email: {\tt P.K.Townsend@damtp.cam.ac.uk}} \\

\end{center}

\vskip 0.5cm

\begin{center} {\bf ABSTRACT}\\[3ex]
\end{center}

A given field theory action determines a set of field equations but  other actions may  yield
equivalent field equations; if so they are  {\it on-shell} equivalent. They may also be  {\it off-shell} equivalent,
being related by the elimination of auxiliary fields or by local field redefinitions, but this is not guaranteed,
as we demonstrate by consideration of the linearized limit of  3D massive gravity models. Failure to appreciate
this subtlety has led to incorrect conclusions in recent studies of the ``Minimal Massive Gravity'' model.

\end{titlepage}

\newpage
\setcounter{page}{1}
\tableofcontents


\section{Introduction}

A classical field theory is fully specified by a set of field equations but it is usually possible
to  find an action from which the  field equations  may be
derived as the conditions for stationarity.  Obviously, these conditions are unchanged by
a change of either the sign or the  scale of the action but the sign is usually fixed
by the requirement of positive energy; or unitarity of the semi-classical theory, and in this context
the scale corresponds to choice of units for $\hbar$.
As a result, there is usually no ambiguity in the choice of action for an {\it interacting} semi-classical field theory
if actions related by field redefinitions and/or elimination of auxiliary fields are considered equivalent.
Ambiguities in {\it relative} scales and  signs may arise after linearization but these can usually be resolved
by reference to the interactions, or by symmetries  inherited from the interacting theory. In the context
of gauge theories, or gravity, the  requirements of gauge invariance and semi-classical unitarity are
usually sufficient to  eliminate ambiguities.

For these reasons, little attention has been paid to ambiguities arising in the choice of action
for fields subject to a given set of field equations. However, this issue has
become relevant recently in the context of 3D massive gravity models; in particular
Topologically Massive Gravity  (TMG) \cite{Deser:1981wh}  and Minimal  Massive Gravity (MMG)
\cite{Bergshoeff:2014pca}.  This is partly because the field equations of TMG and MMG are third order,
rather than second order, and partly because of the ``third-way  consistency''
of the MMG equations (as reviewed in \cite{Bergshoeff:2015zga}).

A characteristic of gauge/gravity field equations that are third-way consistent is that their off-shell
extension requires an action with auxiliary fields that cannot be consistently eliminated from the action,
even though (by definition of ``auxiliary'') they can be eliminated from the field equations. The MMG model of \cite{Bergshoeff:2014pca},
which is a simple modification of TMG, was the first example. This was followed by a gauge theory example in
which the usual 3D $SU(2)$ Yang-Mills equation is modified in a similar way \cite{Arvanitakis:2015oga}; the action
is then equivalent to the difference of two SU(2) Chern-Simons actions plus a mass term that breaks
$SU(2)\times SU(2)$ to the diagonal $SU(2)$ subgroup, which is an action originally proposed in the context of
multiple M2/D2-brane dynamics \cite{Mukhi:2011jp}.

 A key issue for this paper is what happens to  third-way consistent  theories upon linearization, and the
spin-1 example of \cite{Arvanitakis:2015oga} provides a convenient starting point. Linearization of the field equations
yields a triplet of Maxwell actions, as for the standard $SU(2)$-YM theory, and this suggests that the two quadratic
actions must be equivalent. This is indeed the case, as is shown by a simple local field redefinition.   In this spin-1 case,
no distinction arises between on-shell equivalence and off-shell equivalence.

Turning to the spin-2 case, we reconsider the linearization of MMG about an AdS$_3$ vacuum. The field equations
become equivalent to those of linearized TMG in this vacuum, for a rescaled mass, and the corresponding quadratic actions
are therefore {\it on-shell} equivalent.  If one assumes that there is no {\it off-shell} ambiguity then the quadratic MMG
action must be the quadratic TMG action with the rescaled mass;  this leads to the conclusion of \cite{Alkac:2017vgg} (recently
reiterated in \cite{Alkac:2018eck}) that the unitarity problems of TMG with AdS asymptotics are shared by MMG, thus
contradicting the main result of \cite{Bergshoeff:2014pca}.  The problem with this conclusion is that the premise of
no off-shell ambiguity is, in this case,  false.

The  quadratic action for MMG in its AdS$_3$ vacuum was obtained in \cite{Bergshoeff:2014pca} in a form that leads
directly to first-order equations.  In this form, the Virasoro central charges of the asymptotic conformal symmetry
algebra can be read off from the coefficients of the terms that  survive in the limit of infinite graviton mass (in which limit
the result of Brown and Henneaux \cite{Brown:1986nw} for 3D GR is recovered). This step was dealt with briefly
in \cite{Bergshoeff:2014pca} and one purpose of this paper is to give a detailed derivation. We also explain precisely
how this quadratic action is related to an action that directly yields the third-order linearized MMG equations.

However, it is another equivalent form of the quadratic action for linearized MMG that is most useful to a discussion of the issue of on-shell {\it versus} off-shell equivalence. This action is found by a local field redefinition (defined away from the chiral point) followed by an elimination of variables; it is the sum of  a  linearized Einstein-Hilbert   term and  a  standard first-order action for a single free spin-2 mode in AdS$_3$, with coefficients that depend on the MMG parameters.  This action makes clear how off-shell ambiguities can arise because  the {\it relative sign}  between the two terms in this quadratic action cannot be changed by any local field redefinition. Any two distinct MMG models will be  off-shell  {\it inequivalent} if they differ in this relative sign.

The physical parameter space of semi-classical MMG is two-dimensional, and MMG degenerates
to TMG on a  one-parameter curve in this space. On this `TMG curve'  the
relative sign in the quadratic action is fixed, and this is what leads to the unitarity problem
of TMG in an AdS$_3$ vacuum: the so-called ``bulk/boundary clash''.  However, there is a region in the MMG parameter
space in which the relative sign is opposite to that of TMG, and this is how MMG evades the  ``bulk/boundary clash''.
Within a subregion (which is connected once account is taken of an  equivalence relation in parameter space \cite{Arvanitakis:2014xna})
all  semi-classical unitarity conditions are satisfied \cite{Bergshoeff:2014pca}.

TMG/MMG is not the only pair of  3D gravity theories that become on-shell equivalent in the linearized limit but
remain off-shell inequivalent, nor is it necessary for the background to be AdS.  In our concluding section we discuss another
pair of massive 3D gravity theories that become on-shell equivalent when linearized about their Minkowski vacua but for which
off-shell equivalence, even in this linearized limit, is  {\it a priori} obvious!

\section{The spin-1 case}

The gauge-potential one-form $A$ of an $SU(2)$  Yang-Mills (YM) theory is a 3-vector in the Lie algebra of $SU(2)$.
Its two-form field strength is
\begin{equation}
 dA + \frac12 A\times A  =  F \equiv \frac12 dx^\mu dx^\nu F_{\mu\nu}\, ,
\end{equation}
where we use the cross product notation of 3-vector algebra.  We shall suppose that $A$ has dimensions of mass
in units for which $\hbar=1$ so that $F$ has dimension of mass-squared.  For a 3D background Minkowski spacetime,
with metric $\eta$ and standard Minkowski coordinates, the first-order form of the standard 3D YM
Lagrangian density is
\begin{equation}\label{YML}
{\cal L}_{\rm YM} = G_\mu \cdot \tilde F^\mu - \frac12 \eta^{\mu\nu} G_\mu \cdot G_\nu   \qquad
\left(\tilde F^\mu = \frac12 \varepsilon^{\mu\nu \rho} F_{\nu\rho}\right)\, ,
\end{equation}
where $G_\mu$ is an auxiliary $SU(2)$ triplet of Lorentz vectors (of dimension mass-squared)  and we use the dot product notation
of 3-vector algebra to construct an $SU(2)$ singlet.  Elimination of $G$ from this action,
by means of its field equation $G= \tilde F$, yields the standard second-order YM Lagrangian
density in terms of $\tilde F$, which leads to the standard 3D YM field equation in the form
\begin{equation}
\varepsilon^{\mu\nu\rho} D_\nu \tilde F_\rho  =0\, ,
\end{equation}
where $D$ is the covariant derivative with gauge potential $A$.  This equation implies, as a consequence of the Lie algebra
Jacobi identity and the symmetry of mixed partial derivatives, that
\begin{equation}\label{NI}
D_\mu \left[ \varepsilon^{\mu\nu\rho} D_\nu \tilde F_\rho\right] \equiv 0\, ,
\end{equation}
which is the Noether identity required by gauge invariance of the YM action.

Consider now the following modified YM field equation  \cite{Arvanitakis:2015oga}:
\begin{equation}\label{AST}
\varepsilon^{\mu\nu\rho} D_\nu \tilde F_\rho  + \frac{1}{2m}\varepsilon^{\mu\nu\rho}\tilde F_\nu \times \tilde F_\rho = 0\, ,
\end{equation}
where $m$ is a mass parameter.  In light of the identity (\ref{NI}), consistency requires that
\begin{equation}
D_\mu\left[ \varepsilon^{\mu\nu\rho}\tilde F_\nu \times \tilde F_\rho\right] =0\, ,
\end{equation}
but  this is {\it not} an identity. If it were an identity then we might expect to be able to find an action $I[A]$ from which the modified YM
field equation could be derived by variation\footnote{Although  there is no theorem that guarantees this \cite{Deser:2018mje}. };
as it is not an identity, no such action exists!  Nevertheless, there {\it is} an action involving the auxiliary 3-vector $G$; its
Lagrangian density is \cite{Arvanitakis:2015oga}
\begin{equation}\label{YMLmod}
{\cal L} = {\cal L}_{\rm YM} +
\frac{1}{2m}\varepsilon^{\mu\nu\rho}\left(G_\mu\cdot D_\nu G_\rho + \frac{1}{3m} G_\mu\cdot G_\nu\times G_\rho\right)\, .
\end{equation}
It appears that $G$ is no longer auxiliary, but the field equations of $A$ and $G$ are jointly equivalent to $G=\tilde F$ and
(\ref{AST}).  The attempt to find  an action $I[A]$ by setting $G=\tilde F$ in (\ref{YMLmod}) fails because this equation for $G$
is a linear combination of the field equations found from variation of $G$ and $A$, not the one found from variation of $G$ alone.
This is characteristic of gauge field equations that are ``third-way consistent''.

One might wonder whether the modification to the YM Lagrangian density in (\ref{YMLmod}) could be cancelled by a field
redefinition of the form  $A= A' - (\alpha/m)G$ for some constant $\alpha$. Clearly, one may remove {\it either} the
$G\cdot DG$ term {\it or}
 the $G\cdot G\times G$ term in this way, but one cannot remove both. For example, by choosing $\alpha=1/2$ we arrive
 at the new Lagrangian density
 \begin{equation}\label{YMLmodnew}
{\cal L} = {\cal L}_{\rm YM}  - \frac{1}{12m^2} \varepsilon^{\mu\nu\rho}G_\mu\cdot G_\nu\times G_\rho\, .
\end{equation}
The field equation for $G$ is now algebraic, but still non-linear; it can be solved by an infinite series in powers of
$\tilde F/m^2$ \cite{Mukhi:2011jp} but there is no guarantee of convergence.

The above discussion is easily generalized to other gauge groups ${\cal G}$, and as the choice ${\cal G}=U(1)$ is
directly relevant to linearization of the ${\cal G}=SU(2)$ choice, we shall now focus on this. It is obvious that the modified
YM field equation of (\ref{AST}) degenerates to the 3D Maxwell equation for ${\cal G}=U(1)$ but this still leaves open
the possibility of inequivalent actions, and one might expect to find a new non-standard action from the ${\cal G}=U(1)$
variant of (\ref{YMLmod}). However,  the alternative Lagrangian density (\ref{YMLmodnew}) makes it clear that this does not
happen because the cubic term in $G$ is absent for ${\cal G}=U(1)$; the candidate for a new 3D Maxwell action is related to the standard action by a local field redefinition.
In this linearized spin-1 case,  on-shell equivalence implies off-shell equivalence.

\section{Spin-2 in  AdS$_3$}
\label{sec:MMG}
\newcommand{\cG}{{\cal G}}

A complication of the spin-2 case is that the vacuum spacetime must now be found as a solution of the
field equations.  Here we shall be concerned with massive gravity models that admit an AdS$_3$ vacuum.
These include TMG and MMG, and other massive gravity theories that we shall discuss in our concluding section.
We shall see that linearization of TMG and MMG about an AdS$_3$ vacuum leads to massive spin-2
field equations that are  equivalent except for the dependence of the mass parameter on the parameters
that define the TMG and MMG theories. This is to be expected from the fact that MMG involves an additional parameter.
However, we shall also see that the higher-dimensional parameter space  of MMG leads to an off-shell inequivalence
within this parameter space, which allows MMG to be unitary,  at least at the semi-classical level,  even though TMG is not.

\subsection{TMG and MMG actions}

In first-order (Chern-Simons--like) formulation, the TMG Lagrangian can be written in terms of three
Lorentz-vector valued one-form fields: the dreibein $e^a$, a dualized spin-connection $ \omega^a$ and an auxiliary field $h^a$, for $a=0,1,2$.
Using the dot and cross notation for contractions with the invariant bilinear form ($\eta_{ab}$) and structure constants ($\epsilon_{abc}$) of $so(2,1)$, following \cite{Bergshoeff:2014pca}, we may write the Riemann curvature two-form and torsion two-form as, respectively,
\begin{equation}
R(\omega) = d\omega + \frac12 \omega \times \omega\, , \qquad T(\omega) = de + \omega\times e\, ,
\end{equation}
and the  TMG Lagrangian 3-form as
\begin{equation}\label{Ltmg}
L_{\textsc{tmg}} = - \sigma e \cdot R(\omega) + \frac{\Lambda_0}{6} e \cdot e \times e + h \cdot T(\omega)  + \frac{1}{\mu} L_{\textsc{lcs}} (\omega)\,,
\end{equation}
where $\sigma$ is a sign\footnote{One may allow $\sigma$ to be any real number, and this  simplifies the description of the MMG parameter space \cite{Arvanitakis:2014xna}, but here we follow \cite{Bergshoeff:2014pca}.},  $\Lambda_0$ is a `cosmological parameter',  $\mu$ a mass parameter, and $L_{\textsc{lcs}}$ is the Lorentz-Chern-Simons 3-form:
\begin{equation}
L_{\textsc{lcs}}(\omega) = \frac12\omega \cdot \left(d\omega + \frac{1}{3} \omega \times \omega \right)\,.
\end{equation}
As this term breaks parity,  we may choose $\mu>0$ without loss of generality.

After elimination of the auxiliary field $h$ and the dualized spin connection $\omega$ by their field equations,  the TMG action  can be written in terms of the metric alone. In this form it is the sum of an Einstein-Hilbert term and a Lorentz-Chern-Simons term \cite{Deser:1981wh}; the first of these has a coefficient inversely proportional to the 3D Newton constant $G_3$, which has dimensions of inverse mass.
The cosmological parameter $\Lambda_0$ is, for TMG,  the cosmological constant $\Lambda$ in a maximally symmetric background.
Here we shall be interested in the AdS$_3$ case, for which $\Lambda= -1/\ell^2$, where $\ell$ is the AdS radius of curvature.
The semi-classical limit is one for which
\begin{equation}
\frac{\ell}{\hbar G_3} \to \infty\, , \qquad \mu G_3 \to 0\, , \quad {\rm for\ fixed}\quad \frac{\mu\ell}{\hbar}\, .
\end{equation}
The parameter space of semi-classical TMG is therefore one-dimensional, and it is parametrized by the dimensionless parameter
$\mu\ell/\hbar$ \cite{Li:2008dq}.

MMG is defined by the following simple modification of the TMG Lagrangian 3-form:
\begin{equation}\label{Lmmg}
L_{\textsc{mmg}} = L_{\textsc{tmg}} + \frac{\alpha}{2} e\cdot h \times h \,,
\end{equation}
where $\alpha$ is a new dimensionless parameter. The parameter-space of semi-classical MMG is therefore two-dimensional
and MMG degenerates to TMG on the one-dimensional subspace defined by $\alpha=0$.  For $\alpha\ne0$ it is still true that $h$
and $\omega$ can be eliminated from the field equations, which can therefore be written in terms of the metric alone, but the
auxiliary field $h$ can no longer be eliminated from the {\it action}. To understand this unusual state of affairs it is convenient to
express the action in terms of the new (dualized) connection
\begin{equation}
\Omega = \omega + \alpha h\, .
\end{equation}
The action is slightly more complicated in terms of the connection $\Omega$; it reads\footnote{We assume, as in \cite{Bergshoeff:2014pca}, that $(1+ \alpha \sigma) \neq 0$.}
\begin{align}\label{Lmmg2}
L_{MMG} = & - \sigma e \cdot R(\Omega) + \frac{\Lambda_0}{6} e \cdot e \times e + (1+ \alpha\sigma) h \cdot T(\Omega) - \frac{\alpha}{2}(1+\alpha \sigma) e\cdot h \times h \\
\nonumber &  + \frac{1}{\mu} L_{\textsc{lcs}} (\Omega) - \frac{\alpha}{\mu} h \cdot \left( R(\Omega) - \frac{\alpha}{2} D(\Omega) h + \frac{\alpha^2}{6} h\times h \right)\,.
\end{align}
Here $D(\Omega)$ denotes the covariant derivative with respect to the connection $\Omega$. The field equations are:
\begin{subequations}
\begin{align}
\label{Eeqn} - \sigma R(\Omega) + (1+\alpha\sigma) D(\Omega) h - \frac{\alpha}{2}(1+\alpha \sigma) h\times h
+ \frac{\Lambda_0}{2} e \times e & = 0 \, , \\
\label{Oeqn} - \sigma T(\Omega) + \frac{1}{\mu} R(\Omega) - \frac{\alpha}{\mu} D(\Omega) h +
(1+\alpha \sigma) e\times h + \frac{\alpha^2}{2\mu} h \times h & = 0 \, , \\
\label{Heqn} (1+ \alpha \sigma) T(\Omega) - \frac{\alpha}{\mu} R(\Omega) +
\frac{\alpha^2}{\mu} D(\Omega) h - \alpha(1+\alpha \sigma) e\times h - \frac{\alpha^3}{2\mu} h \times h & = 0 \, .
\end{align}
\end{subequations}
By taking linear combinations of these equations, one can show that they are equivalent to the following simpler set:
\begin{subequations}\label{fulleqs}
\begin{align}
T(\Omega) & = 0 \, ,\\
R(\Omega) + \mu (1+\alpha \sigma)^2 e\times h + \frac{\Lambda_0 \alpha}{2} e \times e & = 0\, , \\
D(\Omega) h + \sigma \mu (1+\alpha \sigma) e\times h - \frac{\alpha}{2} h \times h + \frac{\Lambda_0 }{2} e \times e & = 0\, .
\end{align}
\end{subequations}
The first of these equations tells us that the connection $\Omega$ is torsionless; it can be solved for  in terms of $de$. The second equation
allows us to solve for  $h$ in terms of $R(\Omega)$ and $e$. Substituting these solutions into the third equation leads to the MMG field equation presented in \cite{Bergshoeff:2014pca}:
\begin{equation}\label{MMGeqn}
\frac{1}{\mu} C_{\mu\nu} + \bar{\sigma} G_{\mu\nu} + \bar{\Lambda}_0 g_{\mu\nu} = - \frac{\gamma}{\mu^2} J_{\mu\nu}\,,
\end{equation}
where $C_{\mu\nu}$ is the Cotton tensor, $G_{\mu\nu}$ the Einstein tensor and $J_{\mu\nu}$ a curvature squared symmetric tensor
given by
\begin{equation}
J_{\mu\nu} = \frac{1}{2 \det g} \varepsilon_{\mu}{}^{\rho \sigma} \varepsilon_{\nu}{}^{\tau \eta} S_{\rho \tau} S_{\sigma  \eta}\,,
\qquad  S_{\mu\nu} \equiv  R_{\mu\nu} - \frac14 g_{\mu\nu}R\, .
\end{equation}
The coefficients appearing in the MMG field equations \eqref{MMGeqn} are related to the coefficients of the Lagrangian 3-form
\eqref{Lmmg} by
\begin{align}
\bar{\sigma} & = \sigma + \alpha \left[ 1+ \frac{\alpha \Lambda_0/\mu^2}{2(1+\sigma\alpha)^2} \right]\,, & \gamma  & = -\frac{\alpha}{(1+\sigma \alpha)^2}\,, \\
\bar{\Lambda}_0 & = \Lambda_0 \left[1 + \sigma \alpha - \frac{\alpha^3 \Lambda_0/ \mu^2}{4 (1+\sigma \alpha)^2} \right]\,. \nonumber
\end{align}

These manipulations are fine at the level of field equations, but they cannot be used to obtain an action $I[g]$ for which variation with respect to the metric $g$ yields the MMG metric equation (\ref{MMGeqn}).  The reason for this is that a linear combination of all field equations had to be used when solving for $h$ and $\Omega$, so back-substitution in the action is not legitimate; it leads to an {\it inequivalent} action and  a corresponding field equation that is inequivalent to \eqref{MMGeqn}.

\subsection{AdS vacuum and Linearization}

For an  AdS$_3$ vacuum solution of the MMG field equations (\ref{fulleqs}) , we have
\begin{equation}
R(\Omega) = \frac{\Lambda}{2} e \times e \,,  \qquad h = C \mu e\,,
\end{equation}
where $\Lambda$ is the cosmological constant and $C$ a dimensionless constant. These constants
are related to each other and the parameters of the action by
\begin{equation}\label{Cdef}
C = - \frac{(\Lambda + \alpha \Lambda_0)/\mu^2}{2(1 + \alpha \sigma)^2}
\end{equation}
and
\begin{equation}\label{Lambdaeqn}
(\Lambda_0 - \sigma \Lambda)/\mu^2 - \alpha (1+\sigma \alpha) C^2 = 0\,.
\end{equation}

Let $e=\bar e$ be a given AdS$_3$ vacuum solution\footnote{There may be none, one or two, depending on the choice of parameters.}
and $\Omega= \bar\Omega$ the corresponding dual spin-connecton 1-form.
We expand about  this background by setting
\begin{equation}
e = \bar{e} + k \,, \qquad \Omega = \bar\Omega + v\,, \qquad h = C \mu (\bar{e} + k) + p\,.
\end{equation}
The linearized field equations may now be found by expanding the full field equations (\ref{fulleqs})  to first order in the
 perturbation one-forms $(k,v,p)$.  We may also arrive at these linearized equations by first expanding the
 Lagrangian 3-form of  \eqref{Lmmg2} to second order; using \eqref{Cdef} and \eqref{Lambdaeqn}, we find that
 \begin{align}\label{L2}
L^{(2)}_{MMG} = & - (\sigma + \alpha C) \left[k \cdot \bar{D} v + \frac12 \bar{e} \cdot v \times v - \frac{\Lambda}{2} \bar{e} \cdot k \times k \right] \nonumber \\
& - \frac{\Lambda}{\mu} \left[\bar{e} \cdot v \times k + \frac12 k \cdot \bar{D} k - \alpha \bar{e} \cdot k \times p \right] \\
& + (1+\alpha\sigma + \alpha^2 C) \left[p \cdot \bar{D}k + \bar{e} \cdot v \times p - \frac{\alpha}{2} \bar{e} \cdot p \times p \right]  \nonumber \\
& + \frac{1}{\mu} \left[ \left(\frac12 v - \alpha p \right)\cdot \bar{D}v + \frac{\alpha^2}{2} p \cdot \bar{D} p \right]\, ,  \nonumber
\end{align}
where $\bar{D}$ is the covariant derivative with respect to the background spin-connection $\bar{\Omega}$.

The linearized  MMG field equations now follow by variation of the quadratic MMG action with respect to $(k,v,p)$; the resulting equations are jointly  equivalent to
\begin{subequations}\label{lineqs}
\begin{align}
\bar{D} k + \bar{e} \times v & = 0\,, \label{lineq1} \\
\bar{D} v - \Lambda \bar{e} \times k & = - \mu (1+\alpha \sigma)^2 \bar{e} \times p \,, \label{lineq2} \\
\bar{D} p + M \bar{e} \times p & = 0\, , \label{lineq3}
\end{align}
\end{subequations}
where $M$ (the mass of the spin-2 mode) is given by
\begin{equation}
M = \mu(\sigma (1+ \alpha\sigma)- \alpha C)\, .
\end{equation}
These equations should be equivalent to those found by linearization of the third-order metric field equation (\ref{MMGeqn}). To verify this, we first observe that equations (\ref{lineqs}) imply the constraints
\begin{equation}\label{constraints}
\bar{e} \cdot k = \bar{e} \cdot v = \bar{e} \cdot p =0\,.
\end{equation}
If we define
\begin{equation}
k_{\mu\nu} \equiv k_{\mu}{}^a \bar{e}_{\nu}{}^b \eta_{ab}\, ,
\end{equation}
and likewise for the other fields, then the constraints state that the two-tensor fields $(k,v,p)$ are symmetric.
We may solve equations \eqref{lineq1} and \eqref{lineq2} for the symmetric two-tensors $v$ and $p$:
\begin{equation}\label{vpsol}
v_{\mu\nu} = \det(\bar{e})^{-1} \epsilon_{\nu}{}^{\alpha \beta} \bar{\nabla}_\alpha k_{\beta \mu}\,, \qquad p_{\mu\nu} = \frac{2}{\mu(1+\alpha \sigma)^2} \left( \cG_{\mu\nu}(k) - \frac12 \bar{g}_{\mu\nu} \, \cG^\lambda_\lambda(k) \right)\, ,
\end{equation}
where $\cG_{\mu\nu}(k) $ is the linearized Einstein tensor and $\bar{g}_{\mu\nu} = \bar{e}_{\mu}{}^a \bar{e}_{\nu}{}^b \eta_{ab}$ is the background AdS$_3$ metric.
Equation \eqref{lineq3} then becomes
\begin{equation}\label{linmmgeqn}
\epsilon_{\mu}{}^{\alpha \beta} \bar{\nabla}_{\alpha} \cG_{\beta\nu}(k) + M \cG_{\mu\nu}(k) = 0\,.
\end{equation}
This is indeed equivalent to the equation that one obtains from direct linearization of \eqref{MMGeqn}. It is also equivalent to the linearized TMG equation, albeit with a different value for the mass of the spin-2 mode. It is therefore tempting to suppose that  the quadratic action of
linearized MMG  must be equivalent to the quadratic action of linearized TMG.  This was the premise of \cite{Alkac:2017vgg}, which led to the conclusion that the known unitarity problems of TMG persist in MMG.  We  show here that this reasoning is mistaken because the quadratic action of linearized MMG is {\it inequivalent} to the quadratic action of TMG.

\subsection{MMG {\it versus} TMG}
\label{subsec:vs}

For any massive 3D gravity model with  AdS asymptotics,  semi-classical unitarity requires positive energy of the bulk spin-2 modes
{\it and} positive Virasoro central charges for the asymptotic conformal algebra. For TMG it is not possible to satisfy both these conditions
simultaneously. The standard ``wrong sign'' choice for the Einstein Hilbert term in the standard TMG action ensures that the spin-2 mode is physical but this comes at the cost of positivity of the central charges; at least one must be negative. Changing the overall sign
of the action (thereby restoring the ``right sign'' for the Einstein-Hilbert term) will allow both central charges to be positive but this now
comes at the cost of negative energy for the bulk spin-2 mode. The main result of \cite{Bergshoeff:2014pca} is that this ``bulk/boundary clash''
 is resolved by MMG; we shall re-investigate this claim in a way that clarifies its relation to the issue of on-shell versus off-shell
 equivalence.

 In what follows, we shall assume that the overall sign of the action has been chosen such that the Virasoro central charges are positive.
 We introduce a new set of one-form fields $(\tilde k,\tilde v,p)$ to replace the one-form fields $(k,v,p)$ of (\ref{L2}) by setting
\begin{subequations}
\begin{align}
k & = \frac12 (\sqrt{\lambda_-} - \sqrt{- \lambda_+} ) \tilde{k} + \frac{\ell}{2} (\sqrt{\lambda_-} + \sqrt{- \lambda_+} ) \tilde{v} + \frac{1}{\mu(1-2C)} p \\
v & = \frac{1}{2\ell} (\sqrt{\lambda_-} + \sqrt{- \lambda_+} ) \tilde{k} + \frac12 (\sqrt{\lambda_-} - \sqrt{- \lambda_+} ) \tilde{v} + \frac{M}{\mu(1-2C)} p
\end{align}
\end{subequations}
where
\begin{equation}
\lambda_{\pm} = 1 \mp (\sigma + \alpha C) \mu \ell \,.
\end{equation}
This is an invertible field redefinition provided that
\begin{enumerate}
\item $\mp \lambda_{\pm} > 0$. As we explain in the following section, this is equivalent to positivity of both
Virasoro central charges. We note here that this implies that
\begin{equation}
\lambda_+\lambda_- < 0\, .
\end{equation}

\item $1-2C \neq 0$. From the identity
\begin{equation}
1-2C \equiv \frac{ (M\ell)^2 - 1}{(1+\sigma \alpha)^2 (\mu \ell)^2}\,,
\end{equation}
we see that this condition is  equivalent to $|M\ell| \neq 1$. In other words, the change of variables is defined away from
the ``chiral point''  $|M\ell| = 1$.
\end{enumerate}
In terms of the new set of one-form fields $(\tilde k,\tilde v,p)$,  the Lagrangian 3-form of \eqref{L2} takes the form
\begin{align}\label{mmg2}
L^{(2)}_{MMG} = &  \frac{\lambda_+ \lambda_-}{\mu \ell^2} \left( \tilde{k} \cdot \bar{D} \tilde{v} + \frac12 \bar{e} \cdot \tilde{v} \times \tilde{v} + \frac{1}{2\ell^2} \bar{e} \cdot \tilde{k} \times \tilde{k} \right) \\
& + \frac{1}{2\mu(1-2C)} \left[p \cdot \bar{D} p + M \bar{e} \cdot p \times p \right]\,. \nonumber
\end{align}
Varying $\tilde{v}$ now yields the equation $\bar{D} \tilde{k} + \bar{e} \times \tilde{v} =0$, which can be solved for  $\tilde v$;
the corresponding two-tensor is symmetric and given by  \eqref{vpsol} with $k$ and $v$ replaced by
$\tilde{k}$ and $\tilde{v}$.  Using this solution for $\tilde v$ we arrive at the following equivalent quadratic action for
linearized MMG:
\begin{equation}\label{mmgquad}
L^{(2)}_{MMG} =    \frac{\lambda_+ \lambda_-}{\mu \ell^2} \tilde{k}^{\mu\nu} \cG_{\mu\nu} (\tilde{k}) + \frac{1}{2\mu(1-2C)} \left[p^{\mu\nu} \epsilon_{\mu}{}^{\alpha\beta} \bar{\nabla}_{\alpha} p_{\beta \nu}  + M (p^{\mu\nu} p_{\mu\nu} - p^2) \right]\,.
\end{equation}
This action is the sum of two terms: a second-order action for linearized 3D gravity with metric perturbation $\tilde k$, which contains no gauge-invariant local degrees of freedom, and a first-order action for $p$ that describes a single spin-2 mode of mass $M$. Further field redefinitions can change the magnitudes of the coefficients of these terms but not their signs. Our initial assumption of positive Virasoro central charges
has fixed the sign of the coefficient of the $\tilde k$ term, because it implies that $\lambda_+ \lambda_-<0$, but either sign remains possible
for the coefficient of the other term, and this leads to the possibility of off-shell inequivalence.

For MMG it is possible to choose parameters such that the sign of the first-order action for $p$ is either the same as or opposite to the
sign for the  $\tilde k$ term. The bulk/boundary clash can be resolved only if the signs are the same;  this condition just restricts the
parameter space of MMG but it cannot be satisfied by TMG because TMG is the $\alpha\to 0$ limit of MMG and
\begin{equation}
\mu(1-2C) \stackrel{\alpha \to 0}{=} \mu \left(\sigma - \frac{1}{\mu\ell}\right) \left(\sigma+ \frac{1}{\mu\ell} \right) = - \left. \frac{\lambda_+\lambda_-}{\mu\ell^2} \right|_{\alpha=0} \, .
\end{equation}
What we wish to stress here is that the quadratic action for linearized TMG,  with its opposite signs for the two independent terms in the quadratic Lagrangian 3-form (\ref{mmgquad}),  is {\it inequivalent} to the quadratic action for linearized MMG when its parameters are
chosen such the signs are the same, as required by unitarity.

We have stated  that the Virasoro central charges in the asymptotic conformal symmetry algebra are both positive when
$\mp\lambda_\pm>0$. Although this fact did not play an essential role in the above analysis, it is necessary to know how
the parameters $\lambda_\pm$ are related to the central charges if one wishes to read them off from the quadratic action.
This relation was explained briefly in \cite{Bergshoeff:2014pca}; in the following section we provide a more complete
derivation.


\section{Asymptotic symmetries in CS-like theories}
\label{sec:app}

\newcommand{\ione}{{\tt p}}
\newcommand{\itwo}{{\tt q}}
\newcommand{\ithr}{{\tt r}}
\newcommand{\ifou}{{\tt s}}
\newcommand{\ifiv}{{\tt t}}
\newcommand{\extd}{\mathrm{d}}

The action for MMG belongs to the class of theories with a Chern-Simons--like formulation  \cite{Hohm:2012vh,Bergshoeff:2014bia}. These models can be defined in terms of $so(2,1)$-vector valued one-form fields with a bulk action resembling a Chern-Simons theory; these are now included as special cases.  For a review and Hamiltonian analysis of this type of theory we refer to \cite{Bergshoeff:2014bia,Merbis:2014vja}. Here we first recall some results presented in \cite{Grumiller:2017otl}, where the procedure of computing the asymptotic symmetry algebra in general Chern-Simons--like theories was presented. and we then use this to rederive the central charges in MMG for asymptotically AdS$_3$ boundary conditions and compare with the results of \cite{Bergshoeff:2014pca} and \cite{Alkac:2017vgg}. We also explain how these results determine thermodynamic properties of the BTZ black
hole in the context of a given CS-like theory, in particular MMG.

\subsection{The algebra of asymptotic charges }

CS-like models can be defined in terms of a set of $so(2,1)$-vector valued one-form fields labeled by field-space indices $\ione,\itwo,\ithr$, with an action reminiscent of a CS theory:
\begin{equation}\label{eq:shw20}
	I = \frac{k}{4\pi} \int \Big(g_{\ione\itwo}\, a^\ione \cdot \extd a^\itwo + \frac13\, f_{\ione\itwo\ithr}\, a^\ione \cdot a^\itwo \times a^\ithr \Big)\,.
\end{equation}
Here $g_{\ione\itwo}$ and $f_{\ione\itwo\ithr}$ are a completely symmetric field-space metric and structure constants, respectively and $k$ is the overall coupling constant of the theory. As in section \ref{sec:MMG}, we are suppressing wedge products and using dot and cross notation for the contraction of Lorentz indices with $\eta_{ab}$ and $\epsilon_{abc}$ respectively.

The Chern-Simons--like action is invariant under diffeomorphisms by construction and it was shown in \cite{Grumiller:2017otl} that diffeomorphisms are generated by gauge-like transformations which take the fields $a^\ione \to a^\ione + \delta_\xi a^\ione $ with
\begin{equation}\label{eq:shw21}
	\delta_{\xi} a^\ione = \extd\xi^\ione + f^\ione{}_{\itwo\ithr} (a^\itwo \times \xi^\ithr)\,.
\end{equation}
When $\xi^\ione$ is chosen as
\begin{equation}\label{eq:diffeos}
	\xi^\ione = a_\nu{}^\ione \zeta^\nu\,,
\end{equation}
the transformation \eqref{eq:shw21} generates diffeomorphisms on shell
\begin{equation}\label{eq:CSlikediffeo}
	\delta_{\zeta} a_{\mu}{}^\ione = \zeta^{\nu} \partial_{\mu}a_{\nu}{}^\ione + a_{\nu}{}^\ione \partial_{\mu} \zeta^{\nu} + \ldots \stackrel{{\rm on-shell}}{=} {\cal L}_{\zeta} a_{\mu}{}^\ione\, ,
\end{equation}
where the dots refer to terms which vanish by the equations of motion.

In the presence of boundaries, the constraint function generating bulk diffeomorphisms needs to be improved by a boundary term whose variation reads
\begin{equation}\label{eq:shw22}
	\delta Q[\xi^\ione] =  \frac{k}{2\pi}\, \oint  \big(g_{\ione\itwo}\, \xi^\ione \cdot \delta a_\varphi^\itwo\big)\extd\varphi \,.
\end{equation}
This defines the boundary charge of a diffeomorphism parameterized by \eqref{eq:diffeos}.

In order to find the asymptotic symmetry algebra in the general CS-like theories we first specify the boundary conditions for our fields $a^\ione$. They have to solve the field equations (at least asymptotically) and they should come equipped with the specification of what is allowed to fluctuate on the boundary and what is kept fixed;  i.e., which components of the fields carry state-dependent information.

Then we determine the transformations \eqref{eq:shw21} with gauge parameter \eqref{eq:diffeos} that preserve the boundary conditions, up to the transformation of state-dependent functions. In other words, on the left hand side of \eqref{eq:shw21} we specify which components of the fields are allowed to fluctuate. Then we find the asymptotic gauge parameters $\xi^\ione$ by solving for the right hand side of \eqref{eq:shw21}.

After having found the gauge parameters which preserve \eqref{eq:shw21}, the consistency of the boundary conditions can be checked by inserting the result for the gauge parameter into the variation of the charges \eqref{eq:shw22}. This should be finite on the boundary, integrable and conserved. Once these conditions are met, the Poisson brackets will solely receive contributions from the boundary charges on-shell and reduce to the Dirac bracket algebra of boundary charges \cite{Grumiller:2017otl}
\begin{equation}\label{eq:shw23}
	\{ Q[\xi^\ione],Q[\eta^\itwo] \}^* = - \delta_{\eta} Q[\xi^\ione] =  \frac{k}{\pi}\, \oint \extd\varphi \; \tr\left(g_{\ione\itwo}\, \xi^\ione \cdot \delta_{\eta} a_{\varphi}{}^\itwo \right)
\end{equation}
Imposing boundary conditions on $a_{\varphi}{}^\ione$ suffices to determine the asymptotic symmetry algebra. The conditions on the radial component of the fields can be derived by solving the field equations asymptotically.  The time components of the fields can then be found by demanding the boundary conditions on $a_{\varphi}{}^\itwo$ to be conserved under time evolution.

\subsection{Asymptotically AdS$_3$ boundary conditions in MMG}

We will now investigate the asymptotic symmetry algebra for MMG when choosing asymptotically AdS$_3$ boundary conditions. These boundary conditions can be formulated by expanding the metric in Fefferman-Graham gauge, which in three dimensions leads to the Ba\~nados metrics \cite{Banados:1998gg}
\begin{align}\label{banados}
ds^2 = & \; dr^2 - \ell^2\left(e^{r/\ell} dx^+ - e^{-r/\ell} \cL^-(x^- ) dx^-\right) \left(e^{r/\ell} dx^- - e^{-r/\ell} \cL^+(x^+ ) dx^+\right)\, ,
\end{align}
where $x^\pm = t \pm \varphi$.
We formulate our boundary conditions in terms of the dreibein in a suitable local Lorentz gauge. In terms of the generators $T^a$ of the 3D Lorentz algebra $so(2,1)$ we choose
\begin{subequations}\label{eq:dreibein}
\begin{align}
e_\varphi & = - \frac{\ell}{2} e^{-r/\ell}(\cL^+ - \cL^-) T^0 + \frac{\ell}{2} \left( 2 e^{r/\ell} +  e^{-r/\ell} (\cL^+ + \cL^-)\right) T^1 \,, \label{ephi} \\
e_t & = \frac{\ell}{2} \left( 2 e^{r/\ell} -  e^{-r/\ell} (\cL^+ + \cL^-)\right) T^0 + \frac{\ell}{2} e^{-r/\ell}(\cL^+ - \cL^-) T^1  \,, \label{et} \\
e_r & = T^2 \,.
\end{align}
\end{subequations}
The functions $\cL^\pm$ carry state dependent information and are allowed to fluctuate on the boundary. We wish to find the asymptotic symmetry algebra of diffeomorphisms which preserve this form of the dreibein, up to
$\cL^\pm \to \cL^\pm + \delta_\xi \cL^\pm$. First, we need to solve the constraint equations
\begin{equation}
g_{\ione\itwo}\, \extd a^\itwo + \frac12\, f_{\ione\itwo\ithr}\,  a^\itwo \times  a^\ithr = 0\,,
\label{eq:CSlikeeom}
\end{equation}
where $g_{\ione\itwo}$ and $f_{\ione \itwo \ithr}$ are such that \eqref{eq:shw20} gives the MMG action \eqref{Lmmg}. The solution is given as
\begin{equation}\label{vacsol}
\omega = \Omega - \alpha h \,, \qquad h = C \mu e\,,
\end{equation}
with
\begin{subequations}
\begin{align}
\Omega_\varphi & = \frac{1}{2} \left( - 2 e^{r/\ell} +  e^{-r/\ell} (\cL^+ + \cL^-)\right) T^0 - \frac{1}{2} e^{-r/\ell}(\cL^+ - \cL^-) T^1   \,, \label{Omegaphi} \\
\Omega_t & = \frac{1}{2} e^{-r/\ell}(\cL^+ - \cL^-) T^0 - \frac{1}{2} \left( 2 e^{r/\ell} +  e^{-r/\ell} (\cL^+ + \cL^-)\right) T^1   \,,\label{Omegat} \\
\Omega_r & = 0 \,.
\end{align}
\end{subequations}

We are now ready to compute the transformations \eqref{eq:shw21} on the fields. We are assisted in this process by the secondary constraint of MMG, which reads
\begin{equation}
e \cdot h = 0\,.
\end{equation}
This implies for gauge parameters $\xi^{\ione} = a^\ione_\mu \zeta^\mu$ that
\begin{equation}
e \cdot \xi^h = h \cdot \xi^e\,,
\end{equation}
and hence, by \eqref{vacsol}, that $\xi^h = C \mu \xi^e$. The diffeomorphisms preserving the form of \eqref{eq:dreibein} are given by gauge parameters $\xi^e$ and $\xi^\omega = \xi^\Omega - \alpha C \mu \xi^e$ expressed in terms of two arbitrary functions $f^\pm(x^\pm)$
\begin{align}\label{xie}
\xi^e  = &  \, \frac{\ell}{2} e^{-r/\ell} \left( f^+ ( e^{2r/\ell} - \cL^+ ) + f^- (e^{2r/\ell} - \cL^-) + \frac12(f^{+}{}'' +  f^{-}{}'' ) \right)T^0   \\
& + \, \frac{\ell}{2} e^{-r/\ell} \left( f^+ ( e^{2r/\ell} + \cL^+ ) - f^- (e^{2r/\ell} + \cL^-) - \frac12(f^{+}{}'' -  f^{-}{}'' ) \right)T^1  \nonumber \\
& - \frac12(f^+{}' + f^-{}' ) T^2 \nonumber
\end{align}
and
\begin{align}\label{xiomega}
\xi^\Omega & =  \, - \frac{\ell}{2} e^{-r/\ell} \left( f^+  ( e^{2r/\ell} - \cL^+ ) - f^- (e^{2r/\ell} - \cL^-) + \frac12(f^{+}{}'' -  f^{-}{}'' ) \right)T^0   \\
& - \, \frac{\ell}{2} e^{-r/\ell} \left( f^+ ( e^{2r/\ell} + \cL^+ ) + f^- (e^{2r/\ell} + \cL^-) - \frac12(f^{+}{}'' +  f^{-}{}'' ) \right)T^1  \nonumber \\
& + \frac12(f^+{}' - f^-{}' ) T^2\,. \nonumber
\end{align}
These gauge parameters solve \eqref{eq:shw21} with state dependent functions $\cL^\pm$ transforming as CFT stress tensors
\begin{align}\label{eq:cfttrans}
\delta_\xi \cL^\pm = f^\pm \cL^\pm{}' + 2 f^\pm{}' \cL^{\pm} - \frac12 f^\pm{}'''\,.
\end{align}
The next step is to compute the variation of the charges \eqref{eq:shw22} and check whether it is well-defined, finite and integrable. The result we obtain is all of those things and integrates to
\begin{equation}\label{Q}
Q^\pm[f^\pm] =  \frac{\ell}{8\pi G} \left(\sigma \pm \frac{1}{\mu\ell } + \alpha C \right) \int \extd \varphi \; f^{\pm}(x^\pm) \cL^\pm(x^\pm)\, .
\end{equation}

Finally, using \eqref{eq:shw23} together with the transformation properties of the functions $\cL^\pm$ \eqref{eq:cfttrans}, we find that the asymptotic symmetry algebra is given by two copies of the Virasoro algebra for the Fourier modes of $\cL^\pm$ with central charges
\begin{equation}
c^\pm = \frac{3\ell}{2G}  \left(\sigma \pm \frac{1}{\mu\ell } + \alpha C \right) =  \pm \frac{3\ell}{2G} \frac{\lambda_{\mp}}{\mu\ell}\, .
\end{equation}
This result shows that positivity of both $c^+$ and $c^-$ is equivalent to $\pm \lambda_{\mp} > 0$, as claimed in subsection \ref{subsec:vs}.  It also  agrees with \cite{Bergshoeff:2014pca} but differs from the result of \cite{Alkac:2017vgg},  which is based
on a quadratic  action that is  inequivalent to the quadratic approximation to the non-linear MMG action.

\subsection{BTZ thermodynamics}

For constant $\cL^\pm = \frac{2 G}{\ell} (\ell \mathfrak{m} \pm \mathfrak{j})$ the Ba\~nados solutions \eqref{banados} describe BTZ black holes in Einstein gravity with mass $\mathfrak{m}$ and angular momentum $\mathfrak{j}$. These metrics also solve the MMG field equations, but the mass and angular momentum get extra contributions. Using the results of the last section it is particularly easy to compute the BTZ mass in MMG, which corresponds to the asymptotic charge for time translations. From \eqref{eq:diffeos} we see that the gauge parameter corresponding to a time translation is simply $\xi^\ione = a_t{}^\ione$. By inspection of \eqref{et} and \eqref{xie} one can easily verify that this corresponds to choosing $f^\pm = 1$. The BTZ mass is now readily computed from \eqref{Q} as
\begin{equation}
\ell M_{MMG} = Q^+[f^+=1] + Q^-[f^-=1] = \left( \sigma + \alpha C \right) \ell \mathfrak{m} + \frac{\mathfrak{j}}{\mu\ell}\,.
\end{equation}

Similarly, the angular momentum of the black hole, corresponding to the asymptotic charge associated to the Killing vector $\partial_\varphi$, is easily obtained as:
\begin{equation}
J_{MMG} = Q^+[f^+=1] + Q^-[f^-= -1] = \left( \sigma + \alpha C \right) \mathfrak{j} + \frac{\mathfrak{m}}{\mu}\,.
\end{equation}
The mass and angular momentum satisfy the first law of black hole thermodynamics when the entropy of the BTZ black hole in MMG is given by
\begin{equation}\label{BTZentropy}
S = \frac{2\pi}{4G} \left( (\sigma + \alpha C) r_+ + \frac{1}{\mu\ell} r_- \right) = \frac{\pi}{6 \ell} \left(c^+ (r_+ + r_-) + c^-(r_+ - r_-) \right)\, ,
\end{equation}
where $r_\pm$ are the horizon radii of the BTZ black hole; these are given in terms of $\mathfrak{m}$ and $\mathfrak{j}$ by
\begin{equation}
r_\pm = \sqrt{2G \ell (\ell \mathfrak{m} + \mathfrak{j}) } \pm \sqrt{2G \ell (\ell \mathfrak{m} - \mathfrak{j}) }\, .
\end{equation}

The microscopic Cardy formula for the entropy in the canonical ensemble is (see, e.g.  \cite{Compere:2012jk})
\begin{equation}
S = \frac{\pi^2 \ell}{3} \left( c^+ T^+ + c^- T^- \right) \,,
\end{equation}
where $T^\pm$ are the left and right temperatures.  Identifying these as the temperatures of the outer and inner Killing horizons, which are  $T^\pm = (r_+ \pm r_-)/(2\pi \ell^2)$,  one recovers the
macroscopic entropy formula \eqref{BTZentropy}.

\section{Discussion}

The massive 3D gravity models TMG and MMG both propagate a single massive spin-2 mode. Although they differ in their interactions,
linearization about an AdS$_3$ background yields locally equivalent field equations. Nevertheless, the quadratic actions of linearized
TMG and MMG are inequivalent.  This is possible because these quadratic actions include, for an appropriate basis of fields,
 a 3D linearized Einstein-Hilbert term in addition to an action for the spin-2 mode, and this introduces a relative sign that cannot
 be changed by field redefinitions. Moreover, this relative sign is physically significant because it determines whether
 there will be a concordance or a clash between the twin requirements of positive energy for the massive graviton and positive
 Virasoro central charges for the  asymptotic conformal symmetry algebra, both of which are required for semi-classical unitarity.
 Semi-classical  MMG avoids this clash within a region of its two-parameter space, whereas semi-classical TMG does not,
  and this is possible because the linearized action of  MMG is inequivalent to the linearized action of TMG.

 In other words,   the on-shell equivalence of linearized TMG and NMG does not imply an off-shell equivalence of the quadratic
 approximations to the TMG and MMG actions. This is important because all semi-classical unitarity conditions are
 constraints on the coefficients of the terms in this action, and these coefficients, are determined (up to an overall factor)
 by the respective {\it inequivalent} interactions.

We have spelled this out  in detail here because the distinction between on-shell and off-shell inequivalence  of linearized TMG
and MMG is a subtle one that has been overlooked in other discussions in the literature on these massive 3D gravity models.  However, the distinction is an obvious one  in the context of another pair of  3D massive gravity  theories, even when linearized about a Minkowski vacuum: this pair is ``New Massive Gravity'' (NMG) \cite{Bergshoeff:2009hq}  and  (the third-way consistent) ``Exotic Massive Gravity'' (EMG) \cite{Ozkan:2018cxj}.

The field equations of  NMG and EMG become equivalent when linearized about a Minkowski vacuum: they both propagate a parity
doublet of spin-2 modes. However, the quadratic action for linearized EMG is inequivalent to that of linearized NMG. This is because
the EMG action is parity odd whereas the NMG action is parity even and this distinction survives in the quadratic approximation, even
though the linearized field equations are equivalent. A similar result holds for linearization about an AdS vacuum, with important implications for the  two Virasoro  charges of the asymptotic conformal algebra: they are equal in magnitude for both NMG and EMG but they have the same sign for NMG and opposite  sign for EMG.

\subsection*{Acknowledgements}
We thank Julio Oliva for pointing out some minor sign mistakes in an earlier version of this paper. WM is grateful to St\'ephane Detournay for discussion. WM is supported by the ERC Advanced Grant {\it High-Spin-Grav}  and by FNRS-Belgium (convention FRFC PDR T.1025.14 and convention IISN 4.4503.15). PKT is is partially supported by the STFC consolidated grant ST/P000681/1.


\providecommand{\href}[2]{#2}\begingroup\raggedright\endgroup

\end{document}